\documentclass[leqno]{article}
\usepackage{amsmath,amsfonts,amssymb,amsthm}
\usepackage[letterpaper,top=2cm,bottom=2cm,left=3cm,right=3cm, marginparwidth=2cm, 
]{geometry}

\usepackage{graphicx}
\usepackage{dcolumn}
\usepackage{bm}
\usepackage{amsthm}

\usepackage{hyperref}
\hypersetup{
	colorlinks=true,       
	linkcolor=blue,          
	citecolor=blue,        
	filecolor=magenta,      
	urlcolor=blue           
}

\usepackage{amssymb}
\usepackage{epstopdf}
\usepackage{graphicx}
\usepackage{subfigure}
\usepackage{epsfig}
\usepackage{latexsym, amsmath}
\usepackage[english]{babel}
\usepackage{amsmath, amstext, amsfonts}
\usepackage{float}
\usepackage{appendix}

\def\R{{\mathbb R}}

\def\Q{{\mathbb Q}}
\def\N{{\mathbb N}}
\begin{document}





{\centering\LARGE Non-classical point of view of the Brownian motion generation via Fractional deterministic model\par}
\bigskip

{\centering\large H.E.~Gilardi-Vel\'azquez$^1$ and E.~Campos-Cant\'on$^2$ \par \footnotetext[2]{Corresponding Author}}
{\centering\itshape Divisi\'on de Matem\'aticas Aplicadas, Instituto Potosino de Investigaci\'on Cient\'ifica y Tecnol\'ogica A. C., Camino a la Presa San Jos\'e 2055, Col. Lomas 4 Secci\'on, C.P. 78216, San Luis Potos\'i, S.L.P., M\'exico. $^1$hector.gilardi@ipicyt.edu.mx, $^2$eric.campos@ipicyt.edu.mx\par}
%


\begin{abstract}
In this paper we present a dynamical system to generate Brownian motion based on the Langevin equation without stochastic term and using fractional derivatives, i.e.,
a deterministic Brownian motion model is proposed.  The stochastic process is replaced by considering an additional degree of freedom in the second order Langevin equation. Thus it is transformed into a system of three first order  linear differential equations, additionally $\alpha$-fractional derivative are considered  which allow us obtain better statistical properties. Switching Surfaces are established as a part of fluctuating acceleration.
The final system of three $\alpha$-order  linear differential equations does not contain a stochastic term, so the system generates motion in a deterministic way. Nevertheless, from the time series analysis, we found that   the behavior of the system
exhibits  statistics  properties of Brownian motion, such as, a linear growth in time of the mean square displacement, a Gaussian distribution. Furthermore, we use the detrended fluctuation analysis to prove the Brownian character of this motion.\\

\bf{Fractional Brownian motion;
deterministic Brownian motion;
unstable dissipative systems;
DFA analysis.}
\end{abstract}

\section{\label{sec:Introduction}Introduction}

The Brownian motion study has been developed since Robert Brown studies about fertilization process in flowers \cite{brown}.  One of the first to describe Brownian motion was Thorvald N. Thiele, in 1880, in a paper on the method of least squares. At that time the nature of the 
Brownian motion was uncertain therefore there were many open questions about the particle interactions with their environment.  Until 1900, Louis Bachelier, in his PhD thesis applied Brownian motion to the stock and option market fluctuations \cite{librobrown}. The study of Brownian motion was continued by Albert Einstein, who discussed Brownian motion in his work from the point of view of the molecular kinetic theory of heat\cite{Einstein}. It is worth mentioning that Eistein was unaware of the previous work on the subject and he gave the first mathematical description of a free particle Brownian motion. 
 Later, Smoluchowski \cite{Smoluchowski} brought the solution to the problem and attracted the attention of physicists on the problem.

In 1908, Langevin \cite{langevin} obtained the same result as Einstein, using a macroscopically description based on the Newton’s second law,{\it  i.e.}, the dynamical model is based on second-order differential equation with a stochastic term. He said that his approach is ``infinitely simplest" because it was much simpler than the proposed by Einstein. Since the pioneering work of Langevin to model Brownian motion using a stochastic term in the mathematical model, many papers have been devoted to the description of this phenomenon \cite{Smoluchowski}, where  features of this behavior have been defined. According with Van Hove \cite{VanHove} method Prigogine and Grigolini\cite{Prigogine,Grigolini} ``believed" that Brownian
motion can be  derived from  deterministic Hamiltonian models of classical mechanics. But this theory, Hamiltonian models of classical mechanics and the Liouville equation,   involves  grave difficulties as the enormous degrees of freedom and the long times with respect to the duration process.  However the first deterministic  model of Brownian motion was not proposed until Tref\'an et. al.  \cite {Trefan}.  It is important to mention that two approaches can be distinguished: The former, dynamical models are based on a stochastic term and the second, without a stochastic term, {\it i.e.} a deterministic Brownian motion. 

The idea of deterministic Brownian motion has been discussed in hydrodynamic and chemical reactors with oscillatory behavior, where the dynamic is completely deterministic and it is always refereed as ``microscopic chaos". In this sense Tref\'an {\it et. al.} \cite{Trefan} and Huerta-Cuellar {\it et. al.} 
\cite{huerta} have proposed a Brownian motion generators based on Langevin equation. The idea of Tref\'an {\it et. al.} consist of replacing the stochastic term by a discrete dynamical system which generates pseudo-random numbers; The process drives a Brownian particle and has ``statistical" properties that differ markedly from the standard
assumption of Gaussian statistics due to the discrete dynamical system has a ``U-shaped probability" distribution.  The approach of Huerta-Cuellar {\it et. al.} is in the same spirit that Tref\'an {\it et. al.} but now the stochastic term is controled by the jerky equation, {\it i.e.},  by adding an additional degree of freedom to the Langevin equation it is possible to transform it into a system of three linear differential equations without a stochastic term; this process display ``like- Gaussian probability" distributions of system variables, which differ  from the standard assumption. Based on this approach we use fractional derivatives in this work and get a deterministic Brownian motion with  a greatly improved Gaussian probability distributions of all the variables of the system.

The Langevin equation has been used in many areas, such as modeling the evacuation processes \cite{Kosinski}, photoelectron counting \cite{Wodkiewicz}, analyzing the stock market \cite{Bouchaud}, studying the fluid suspensions \cite{Hinch}, deuteron-cluster dynamics \cite{Takahashi}, protein dynamics \cite{Schluttig}, self organization in complex systems \cite{Fraaije}, etc. For other applications of the Langevin equation in physical chemistry and electrical engineering, one can refer to Ref. \cite{Coffey}. The classical study of the Brownian motion via the Langevin equation is under the hypothesis that the process is a Markov process, {\it  i.e.}, the random forces modeled by the stochastic term are independent so the process does not have memory. Although the Langevin equation plays an important role in many fields, there are still some behaviors such as anomalous diffusion (superdiffusion and subdiffusion), power law, long-range interactions that the Langevin equation can not well describe them. Therefore, various fractional Langevin equations were proposed in Ref.~\cite{Coffey,Kobelev,Mainardi}.  In this way the fractional Langevin equation can capture the aforementioned  features that the Langevin equation can not achieve.

Brownian motion behavior is characterized by specific properties, such as a linear growth in time  of the mean square displacement, Gaussian probability distributions of system variables, an exponential in time decay of the positional autocorrelation function. This last property can be characterized  by the detrended fluctuation analysis  (DFA) which was developed by Peng {\it et al.} \cite{Peng}. So the DFA thecnique characterizes the correlation properties of a signal. In this paper  a deterministic fractional model to generate Brownian motion based on Langevin equation and jerky equation is proposed in the same way that Ref.~\cite{huerta},  its behavior is characterized by time series analysis  via DFA.   
 
This paper is organized as follows: In section 2, some basic concepts of fractional differential equations theory are introduced as differential operators, stability theory in fractional systems and a numerical method to solve fractional systems. In section 3, the  proposed model via jerk equation and fractional derivatives is presented with its stability analysis.  Section 4 contains the numerical results obtained with the proposed model, and the statistical properties of the time series by means of DFA analysis to confirm the Brownian behavior. Also the maximum Lyapunov exponent is computed.  Finally conclusions are drawn in Section 5.

\section{\label{secc:cfrac}Basic concepts in fractional calculus}
Fractional derivatives and integrals are generalizations of integer ones.  Nevertheless, in literature we can find many different definitions for fractional derivatives Ref.~\cite{1,2,3,4} being the Riemann-Liouville and the Caputo definitions the most reported \cite{1}. The fractional derivative of Riemann-Liouville is defined as:
\begin{equation}
D_a^\alpha f(x)=\frac{1}{\Gamma(n-\alpha)}\frac{d^n}{dx^n}\int_a^x\frac{f(t)}{(x-t)^{\alpha-n+1}}dt,
\end{equation}
   
and the Caputo definition is described by:

\begin{equation}
D_0^\alpha f(x)=\frac{1}{\Gamma(n-\alpha)}\int_a^x\frac{f^{(n)}(t)}{(x-t)^{\alpha-n+1}}dt,
\end{equation}
    
with $n=\lceil{\alpha}\rceil$, and $\Gamma$ is the Gamma function which is defined as:

\begin{equation}
\Gamma(z)=\int_0^\infty t^{z-1} e^{-t}dt.
\end{equation}

For instance, in fractional order systems the stability region depends on the derivative order $\alpha$ as it is depicted in  Fig. 1 of  Ref.~\cite{neto}. It is important to note that the stability of an equilibrium point can be controled by means of the derivative order $\alpha$, for example, a saddle hyperbolic equilibrium point of an integer system can be transformed  to a stable equilibrium point by changing the system derivative order  $\alpha$. 
This is an important consideration for designing a mathematical model of Brownian motion because we are interested in unstable dynamics.

A  general commensurate fractional order  time invariant system is described as follows:

\begin{equation}
D^{n_k}_0x(t)=f(t,x(t),D^{n_1}_0x(t),D^{n_2}_0x(t),\cdots,D^{n_{k-1}}_0x(t)),
\label{eq:general}
\end{equation}
subject to initial conditions 
\[ 
x^{(j)}(0)=x_0^{(j)}, \text{ with } j=0,1,\dots, \lceil{n_k}\rceil -1,
\]
where $n_1,n_2,\ldots, n_k\in \Q$,  such that $n_k>n_{k-1}> \cdots >n_1>0$, $n_j-n_{j-1}\leq 1$ for all $j=2,3,\ldots,k$ and $0<n_1 \leq 1$. Let $M$ be the least common multiple of the denominator of $n_1,n_2,\ldots,n_k$ and set $\alpha=1/M$ and $N=Mn_k$. Then accordingly with theorem 8.1 given  in Ref.~\cite{1} the equation (\ref{eq:general}) is equivalent to the following system of equations.

\begin{eqnarray}
D^\alpha_0x_0(t)&=& x_1(t), \nonumber\\
D^\alpha_0x_1(t)&=& x_2(t), \nonumber\\
&\vdots&\\
D^\alpha_0x_{N-2}(t)&=& x_{N-1}(t), \nonumber\\
D^\alpha_0x_{N-1}(t)&=& f(t,x_0(t),x_{n1/\alpha}(t),\cdots,x_{n_{k-1}/\alpha}(t)). \nonumber
\end{eqnarray}
with initial conditions
\[
x_j(0)=\left\{
\begin{array}{ll}
x_0^{(j/M)}& \text{ if } j/M\in \N \cup \{0\}\\
0 & \text{other case.}
\end{array} 
\right.
\]

Furthermore, this linear time invariant system can be expressed in a matrix form as follows:

\begin{equation}
\frac{d^\alpha {\bf x}(t)}{dt^{\alpha}}=A{\bf x},
\label{eq:lineal} 
\end{equation} 

\noindent where ${\bf x}\in \R^n$  is the state vector,  $A \in \R^{n \times n}$ is a linear operator,  and $\alpha$ is the fractional commensurate derivative order $0<\alpha<1$. The stability of this kind of systems is enunciated in as follows:\\

\begin{enumerate}
	\item [-] \emph{\bf Asymptotically stable:} The system (\ref{eq:lineal}) is asymptotically stable if and only if $| arg(\lambda)| >\frac{\alpha\pi}{2}$ for all eigenvalues $(\lambda)$ of matrix $A$. In this case, the solution $x(t) \to 0$ as $t\to \infty$.\\
	
	\item [-] \emph{\bf Stable:} The system (\ref{eq:lineal}) is stable if and only if $| arg(\lambda)| \geq \frac{\alpha\pi}{2}$  for all eigenvalues $(\lambda)$ of matrix $A$ obeying that the critical eigenvalues must satisfy $| arg(\lambda)| = \frac{\alpha\pi}{2}$  and have geometric multiplicity of one.
\end{enumerate}

The interest is to have unstable dynamics in order to generate Brownian motion, thus the system is restricted to have at least one eigenvalue in the unstable region, {\it i.e.}, 
 the system (\ref{eq:lineal}) is unstable if and only if $| arg(\lambda)| < \frac{\alpha\pi}{2}$  for at least one of its eigenvalues $(\lambda)$ of matrix $A$.

Additionally, the system given by \eqref{eq:lineal} with equilibrium point at the origen can be generalized by affine linear system as follows:

\begin{equation}
\frac{d^\alpha {\bf x}(t)}{dt^{\alpha}}=A{\bf x}+B,
\label{eq:affinelineal} 
\end{equation} 
where $B\in \R^n$ is a constant vector and $A \in \R^{n \times n}$ is a nonsingular linear operator. Now the equilibrium point $p\equiv(x_1^*,x_2^*,\cdots,x_N^*)^T=-A^{-1}B$
of a general commensurate fractional order affine linear system (\ref{eq:affinelineal}) with fractional order $0 < \alpha < 1$, is saddle equilibrium point if its eigenvalues $\lambda_1, \lambda_2, \ldots,\lambda_\kappa,\lambda_{\kappa+1},\ldots,\lambda_n $ of its Jacobian matrix evaluated at the equilibrium point fulfill the following condition.

\begin{equation}
\begin{array}{ll}
|arg(\lambda_i)|>\frac{\alpha\pi}{2}& \;\text{ with}\; i=1,2,\ldots,\kappa,\\
|arg(\lambda_i)|<\frac{\alpha\pi}{2}& \;\text{with}\; i=\kappa+1,\kappa+2,\ldots,n.
\end{array}
\label{saddlepoint}
\end{equation}

Note that we are interested in working with unstable systems, {\it i.e.}, systems that do not fulfil the locally asymptotic stable condition.
\begin{equation}
min|arg(\lambda_i)|>\frac{\alpha\pi}{2}, \text{  for}\;  i=1,2,\ldots,n.
\label{critest}
\end{equation}

\subsection{Numerical method to solve fractional differential equations}

There are not methods that can provide  analytically  the exact solution of any fractional differential equation  as those given by integer-order systems, therefore it is necessary to use numerical methods. The Adams-Bashforth-Moulton (ABM) method, a predictor–corrector scheme, was reported
in Ref.~\cite{Diethelmmet} and it is used to obtain the time evolution of fractional systems. The algorithm is a generalization of the classical Adams-Bashforth-Moulton integrator that is well known for the numerical solution of first-order problems as switching systems \cite{neto}.
We now present the method that is well understood and that has been proven to be efficient in many practical applications \cite{Ford,183libro}.

Consider the fractional differential equation that is described by \eqref{eq:general} as follows:

\begin{equation}
\begin{matrix}
D^{\alpha}x(t)=f(t,x(t)),~~~~~~0\leq t\leq T;\\
x^{(k)}(0)=x_0^{(k)},~~~~k=0,1,\dots,n-1.
\end{matrix}
\label{eq:metodo}
\end{equation}

We assume the function $f$ is such that a unique solution exists on some interval
$[0,T]$,  and assume that we are working on a uniform grid $\{ t_j = jh : j = 0,1, \dots ,N \}$ with some integer $N$ and $h = T/N$.

The solution of (\ref{eq:metodo}) is given by an integral equation of
Volterra type as

\begin{equation}
x(t)=\sum_{k=0}^{\lceil \alpha\rceil -1}x^k_0\frac{t^k}{k!}+\frac{1}{\Gamma(\alpha)}\int_0^t(t-z)^{\alpha-1}f(z,x(z))dz,
\end{equation}

\begin{equation}
x_{k+1}=\sum_{k=0}^{\lceil \alpha\rceil -1}x^k_0\frac{t^k}{k!}+\frac{1}{\Gamma(\alpha)}(\sum_{j=0}^ka_{j,k+1}f(t_j,x_j)+a_{k+1,k+1}f(t_{k+1},x_{k+1}^P)),
\end{equation}

\noindent where

\begin{equation}
\begin{array}{l}
a_{j,k+1}=\\ \\
\left\{
\begin{array}{ll}
\displaystyle \frac{h^\alpha}{\alpha(\alpha+1)}(k^{\alpha+1}-(k-\alpha)(k+1)^\alpha), & if ~~ j=0;\\
\displaystyle\frac{h^\alpha}{\alpha(\alpha+1)}((k-j+2)^{\alpha+1}+(k-j)^{\alpha+1} & \\
 -2(k-j+1)^{\alpha+1}),& if ~~ 1\leq j\leq k;\\
\displaystyle\frac{h^\alpha}{\alpha(\alpha+1)}, & if ~~ j=k+1.
\end{array}
\right.
\end{array}
\end{equation}

\noindent With the predictor structure given as follows.

\begin{equation}
x_{k+1}^P=x(0)+\frac{1}{\Gamma(n)}\sum_{j=0}^k b_{j,k+1}f(t_j,x_j),
\end{equation}

\noindent and

\begin{equation}
b_{j,k+1}=\frac{h^\alpha}{\alpha}((k+1-j)^\alpha-(k-j)^\alpha).
\end{equation}

The error of this approximation is given by

\begin{equation}
max_{j=0,1,\dots,N}|x(t_j)-x_h(t_j)|=O(h^p).
\end{equation}

\section{Deterministic model to generate Brownian motion}

The onset of Brownian motion is a suspending particle  in fluids. The motion of this particle occurs due to collisions between molecules of the  fluid, and considering that in each collision  a molecule changes its velocity  in small amount. This fact is because of the suspended particle  under normal conditions  suffers about $10^{21}$ collisions per second, so  the accumulated effect results to be considerable. Each of these collisions is always determined by the last event which is produced by physical interactions in the system. Since it can be thought that each collision produces a kink in the path of the particle, one can not hope to follow the path in any detail, i.e., the details of the path are infinitely fine. Thus the Brownian particle make a fluctuating movement. So stochastic models of Brownian motion follow the average motion of a particle, not a particular path of a particule.

The stochastic theory of Brownian motion of a free particle (in the absence of an external field of force) is generally governed by the Langevin equation. 
\begin{equation}
\begin{array}{l}
\frac{d x}{dt}=v,\\
\frac{d v}{dt}=-\gamma \frac{d x}{dt} +A_f(t),
\end{array}
\label{eq:langevin}
\end{equation}
\noindent where $x$ denotes the particle position and $v$ its velocity. According to this equation, the influence of the surrounding medium on the particle motion can be splitted into two parts. The first term $-\gamma \frac{d x}{dt}$ stands for the friction applied to the particle, it is assumed that the friction term is according with the Stokes’ law which states that the friction force  $6\pi a \sigma v/m$ decelerates a spherical particle of radius $a$ and mass $m$. Hence, the friction coefficient is  given as follows:

\begin{equation}
\gamma=6\pi a \sigma/m,
\end{equation}
 
\noindent where $\sigma$ denotes the viscosity of the surrounding fluid.

The second term  $A_f(t)$ is the fluctuation acceleration which provides a stochastic
character of the motion and depends on the fluctuation force $F_f (t)$ as $A_f (t)=F_f (t)/m$, where $m$ is the particle mass.
 
Two principal assumptions were made about this stochastic term $A_f (t)$ in order to produce Brownian motion:
\begin{enumerate}
\item [-]$A_f (t)$ is independent of $x$ and $v$.
\item [-]$A_f (t)$ varies extremely fast as compared with the variation of $v$.
\end{enumerate}
The latter assumption implies that there exists a time interval $\Delta t$ during which the variations in $v$ are very small. Alternatively,
we may say that though $v(t)$ and  $v(t+\Delta t)$ are expected to differ by a negligible amount, no correlation between
$A_f(t)$  and $A_f(t+\Delta t)$ exists due to it is a stochastic term. 

Based on fractional calculus, various fractional Langevin equations have beeen proposed to generate Brownian motion, see Ref.~\cite{Coffey,Kobelev,Mainardi}. These models differ from the usual Langevin equation by replacing the derivatives with respect to time by the fractional derivative of $\alpha$ order.

\begin{equation}
\frac{d^\alpha v}{dt^\alpha}=-\gamma \frac{d^\alpha x}{dt^\alpha} +A_f(t),
\label{eq:langevinfrac}
\end{equation}

\noindent where $x$ denotes the particle position and $v$ its velocity. 
We can rewrite the fractional Langevin equation (\ref{eq:langevinfrac}) in two fractional differential equations 
by a change of variables,

\begin{equation}
\begin{array}{lll}
D^\alpha_0x&=& v, \\
D^\alpha_0v&=& -\gamma v+ A_f(t), 
\end{array}
\label{sislanfrac}
\end{equation} 

\noindent where $D^\alpha_0$ is the Caputo derivative operator.

In order to generate a deterministic model of  Brownian motion an additional freedom degree is added to the system (\ref{sislanfrac}) in order to avoid the stochastic term by means of replacing the fluctuating acceleration $A_f(t)$. We change the stochastic term $A_f$  for a new variable $z$ which is defined by a third order differential equation as it was reported in Ref.~\cite{huerta}. The proposed variable $z$  acts as fluctuating acceleration, and produces a deterministic dynamical motion without stochastic term but the behavior present the statistics features of Brownian motion as it was showed in previous work \cite{Trefan}. However, in our model the fluctuation acceleration has a direct dependence on the position, velocity and
acceleration due to the jerky equation involved \cite{33huerta}. When a particle is moving in a fluid, friction and collisions with other
particles, existing in the environment, necessarily produces changes in the motion velocity and acceleration; all these changes
are considered in the jerky equation. Without loss of generality, we define our approach based in a unstable dissipative systems (UDS)\cite{34huerta,35huerta}   as follows

\begin{equation}
\begin{array}{lll}
D^{\alpha}{x}&=&v, \\
D^{\alpha}{v}&=&-\gamma v+ z,\\
D^{\alpha}{z}&=&-a_1 x -a_2 v -a_3 z -a_4(x),
\end{array}
\label{eq:modeloprop}
\end{equation}
\noindent where $a_{i }\in \R$ are constant parameters, with $i= 1,2,3$, and $a_4(x)\in \R$ acts as a constant piecewise function, {\it i.e.}, a step function.\\

\begin{figure}
	\centering
		\includegraphics[width=0.70\textwidth]{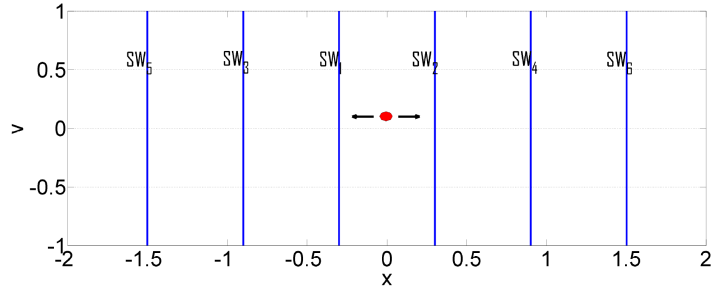}
	\caption{Projection of SW perpendicular on to the plane $(x,v)$ (Blue lines), the red dot depict a Brownian particle that moves along one dimension $x$. The SW delimit each potential region and when the particle cross it represent a potential change on the particle}
	\label{fig:SW}
\end{figure}

The first two equations of  the fractional Langevin equation \eqref{sislanfrac} are derived from thel Langevin equation (\ref{eq:langevin}) with a little change:
the stochastic term is replaced by a deterministic term in the same spirit that \cite{34huerta,35huerta}. Now we construct switching surfaces (SW), see Figure \ref{fig:SW}. Without loss of generality, the SW are defined by  perpendicular planes to the $x$ axis, so domains are defined between these SW's which are considered as edges of each domain.   In case of real systems, SW can be seen
as multi-well potential with short fluctuation escape time where each domain defined by SW's preserve its unstable behavior according to lineal part of system. The parameter $a_4$ is defined as follows

\begin{equation}
a_4(x)=c_1 round(x/c_2),
\end{equation}
with $c_1, c_2\in \R$ are constants. Here, the $round(x)
$ function will be implemented to simplify the SW generation process. The function will be defined as follows:

\begin{equation}\label{ec_round_def}
round(x)=\left\{%
\begin{array}{l}
 \lceil x-1/2 \rceil,  \text{ for } x<0;\\
\lfloor x+1/2 \rfloor, \text{ for } x \geq 0.
\end{array}%
\right.
\end{equation}

\begin{figure}
\centering
\subfigure[]{\includegraphics[width=0.7\textwidth]{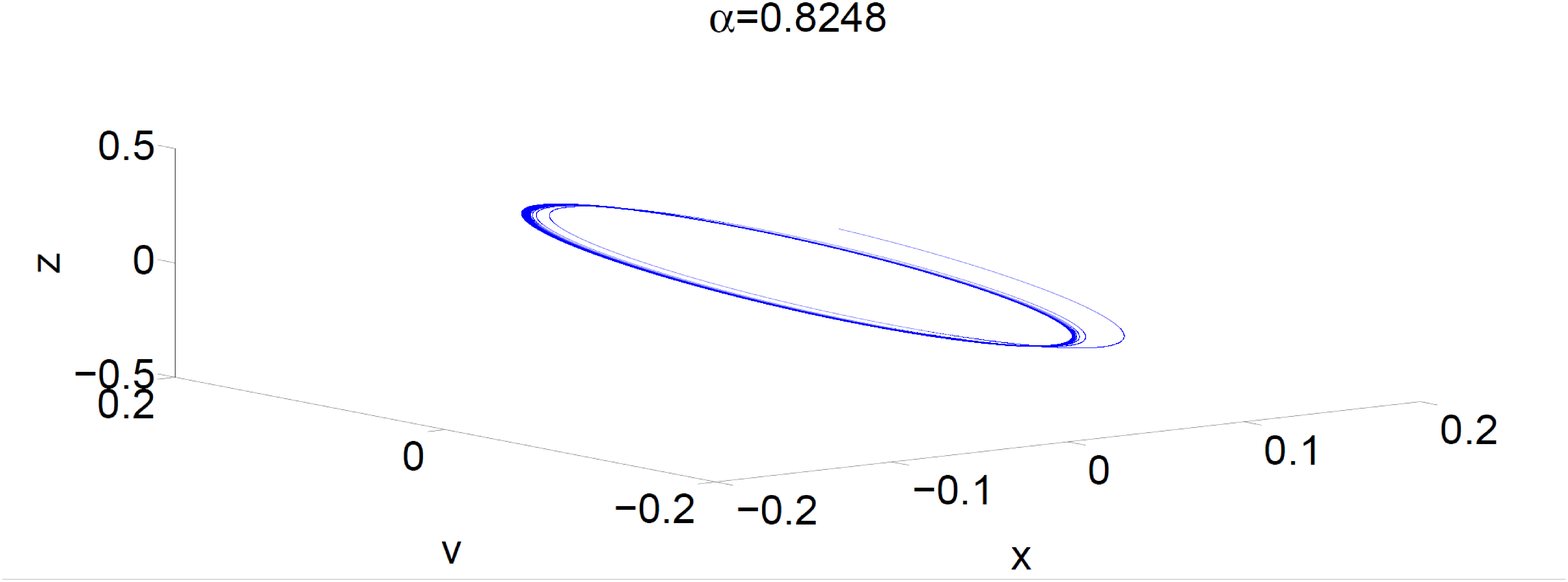}}
\subfigure[]{\includegraphics[width=0.7\textwidth]{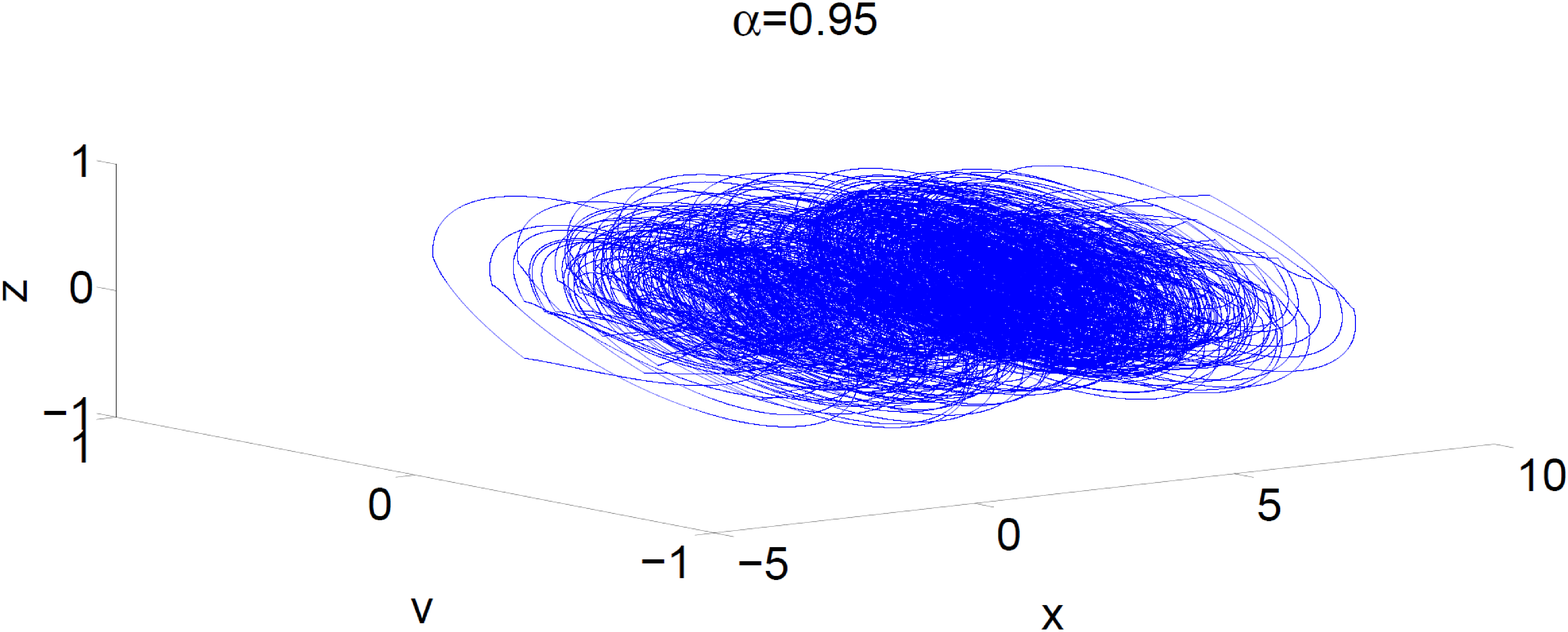}}
\caption{Solutions of the system (\ref{eq:modeloprop}) on phase space for different $\alpha$ values: (a) $\alpha=0.8248$, (b) $\alpha=0.9500$.} \label{fig:difalpha}
\end{figure}
\section{Numerical results}
In this section, we numerically investigate the long term behavior of the solutions of equation  (\ref{eq:modeloprop}) by  considering different derivative orders and the following parameter values: $\gamma=7\times 10^{-5};~~a_1 = 1.5; ~~a_2 = 1.2; ~~a_3 = 0.1;~~ C1 = 0.9$ and $C2 = 0.6$. It is worth to mention that these parameter values are the same to those used  to generate Brownian behavior in Ref.~\cite{huerta} by  considering an integer order system. To study Brownian motion generated by Eq. (\ref{eq:modeloprop}), we fix the parameters values and explore different derivative orders ($\alpha$ values). The stability of fractional order  systems is givern by Eq.(\ref{critest}), so the local behavior near the equilibrium point is determined by the Jacobian of the system (\ref{eq:modeloprop}) which has the following spectrum $\Lambda=\{  \lambda_1= -0.8304, \lambda_2=   0.3651 + 1.2935i, \lambda_3=   0.3651 - 1.2935i\}$. This spectrum determines the critical value of derivative order $\alpha_c\approx 0.8249$ to get the system (\ref{eq:modeloprop}) to be stable,
\begin{equation}
\alpha<\frac{2}{\pi} min|arg(\lambda_i)|\approx 0.8249.
\end{equation}
Accordingly with the aforementioned comments, in order to preserve the system  (\ref{eq:modeloprop}) to be unstable to generate oscillatory behavior, we consider $\alpha>\alpha_c$.  Figure \ref{fig:difalpha}~(a)  shows  a solution of the system  (\ref{eq:modeloprop}) for $\alpha=0.8248$ and initial condition $(x_0,v_0,z_0)^T=(1.0,1.0,1.0)^T$, this derivative order value results in a stable behavior due to $\alpha<\alpha_c$. On the othe hand, Figure \ref{fig:difalpha}~(b)  shows  a solution of the system  (\ref{eq:modeloprop}) for $\alpha=0.95$ and the same inictial condition, but now this derivative order value results in an unstable behavior due to $\alpha>\alpha_c$. Numerical simulations are performed using the  Adams-Bashforth-Moulton algorithm by exploring different $\alpha$ values.

\begin{figure}[h]
	\centering
		\includegraphics[width=1\textwidth]{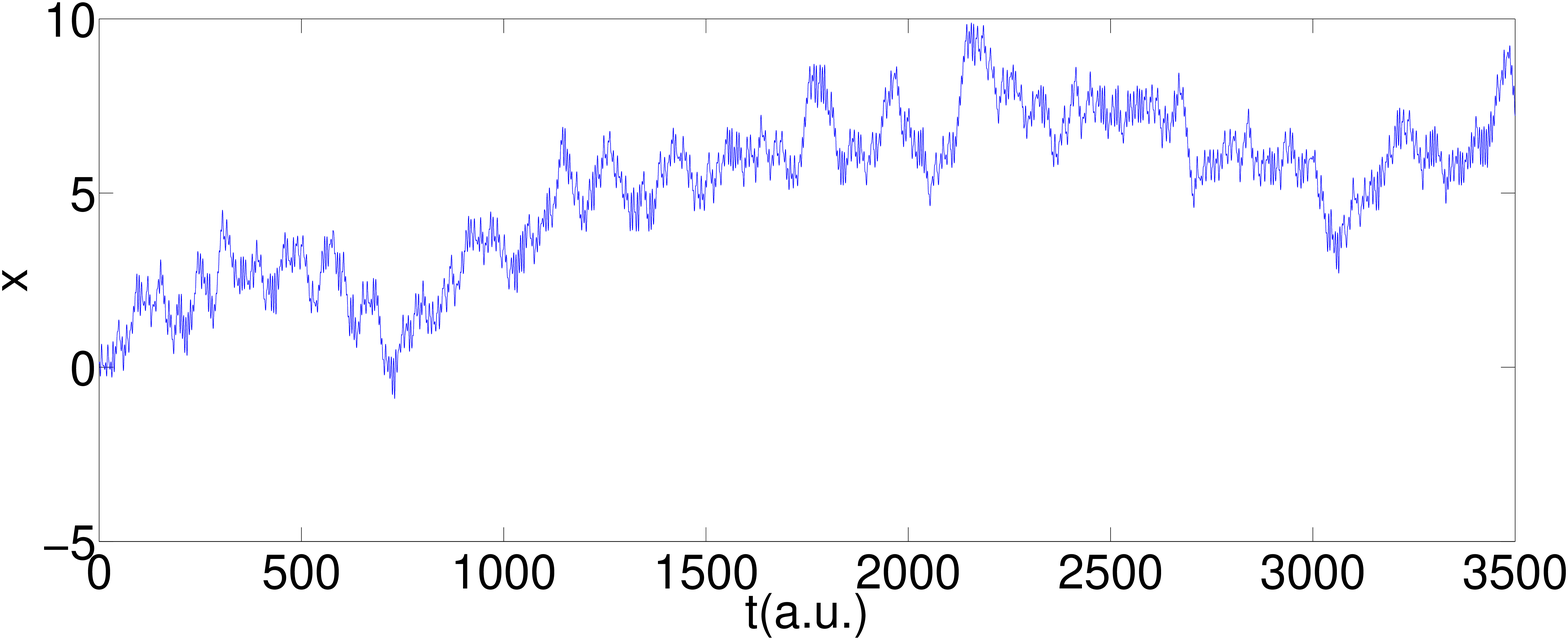}
	\caption{Time series $x$ of deterministic Brownian motion  by the proposed model given by  \eqref{eq:modeloprop} with $\alpha=0.95$.}
	\label{fig:Brown95}
\end{figure}

 Figure \ref{fig:Brown95} shows a time series of a particle position for $\alpha=0.95$ where characteristic behavior of Brownian motion can be clearly seen.  The trajectory of Brownian motion is determined by initial conditions and the parameter values before mentioned.  Since there are many steps with short time duration and few steps with long time duration, the predicted mean square displacement in short times is observed.

\begin{figure}
\centering
\includegraphics[width=0.95\textwidth]{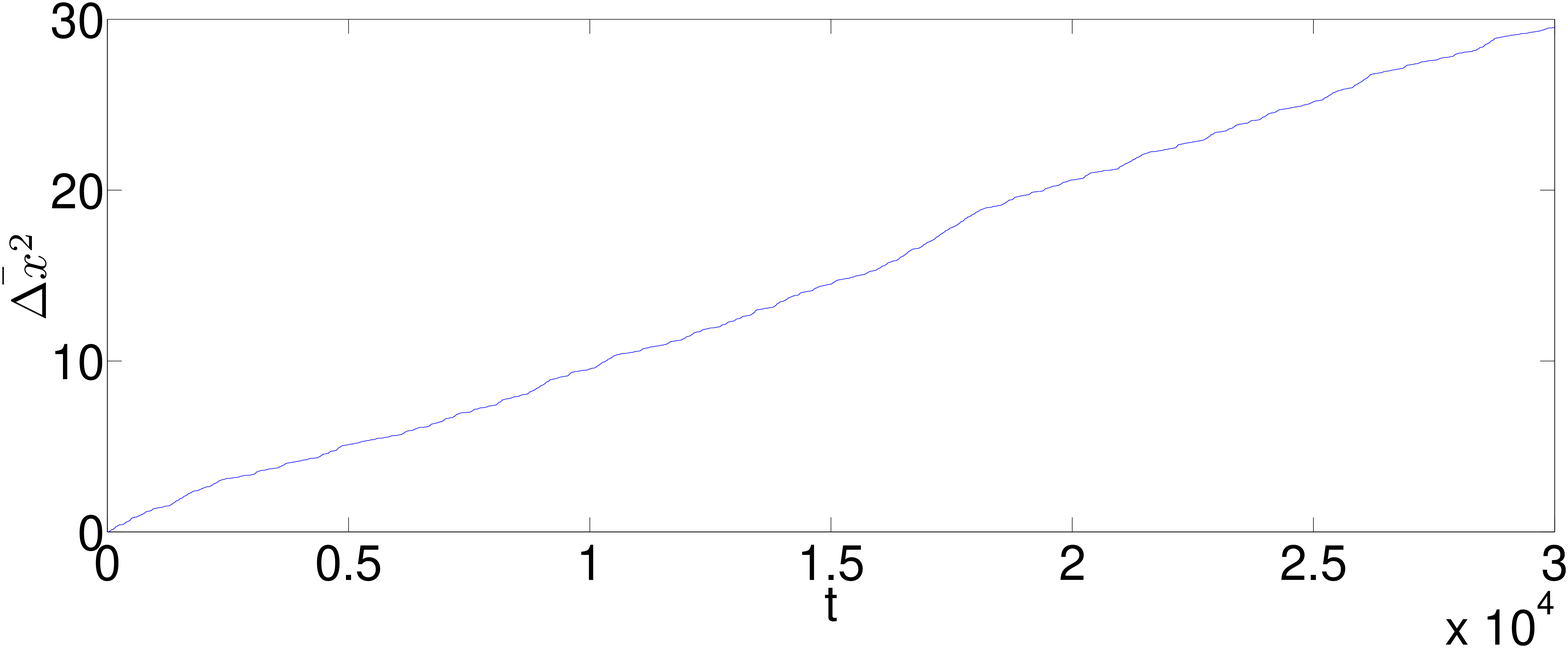}
\includegraphics[width=0.95\textwidth]{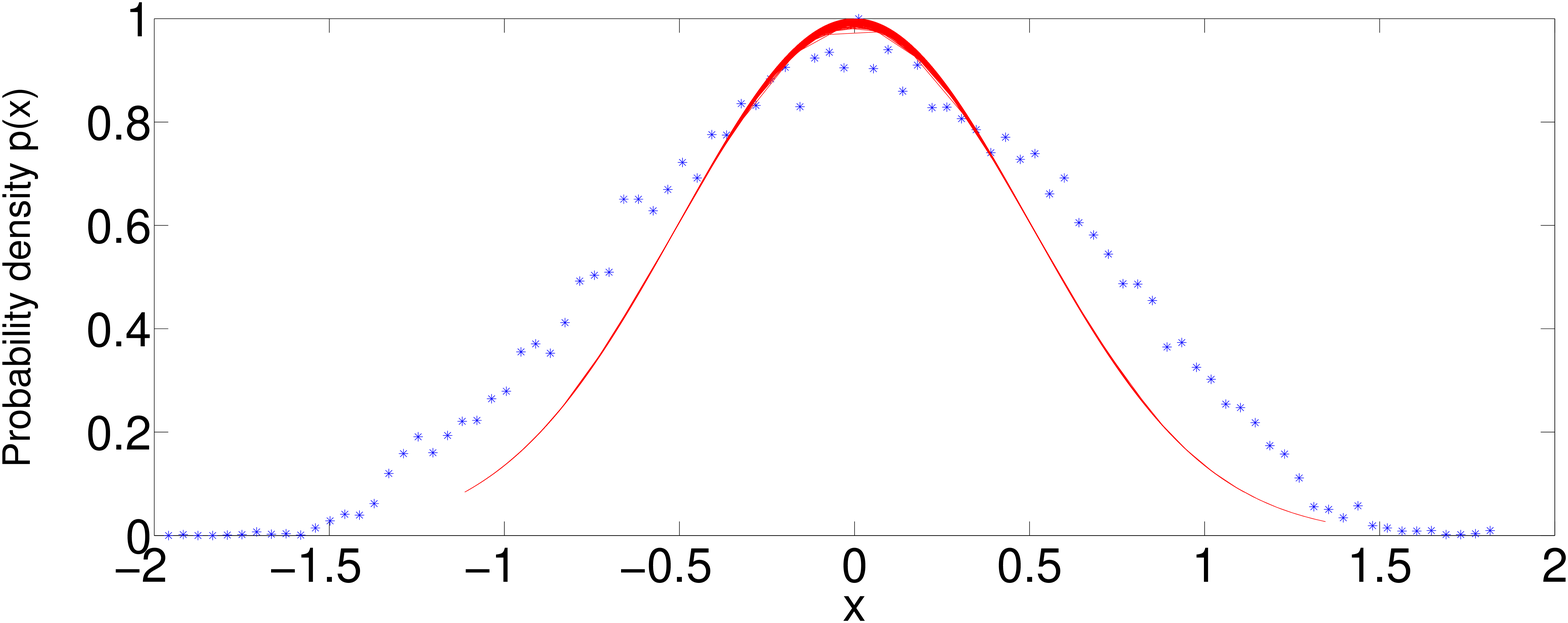}
\includegraphics[width=0.95\textwidth]{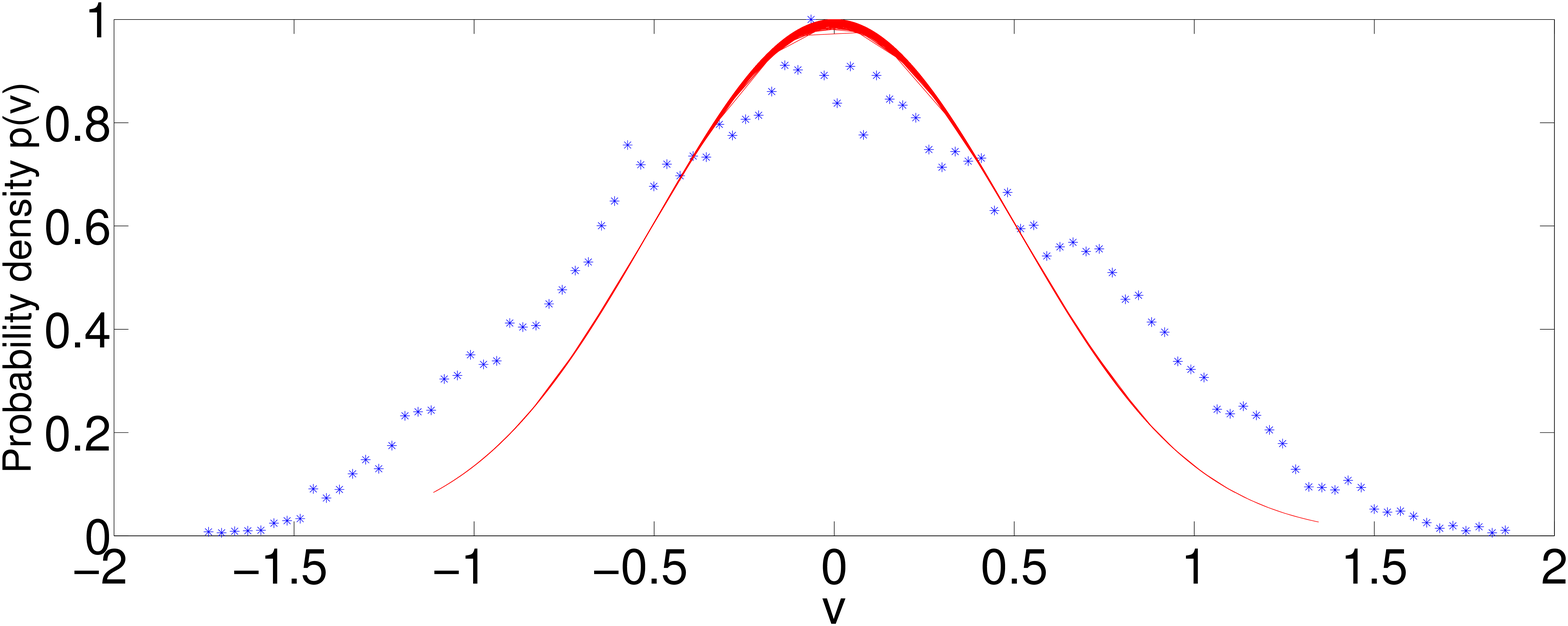}
\includegraphics[width=0.95\textwidth]{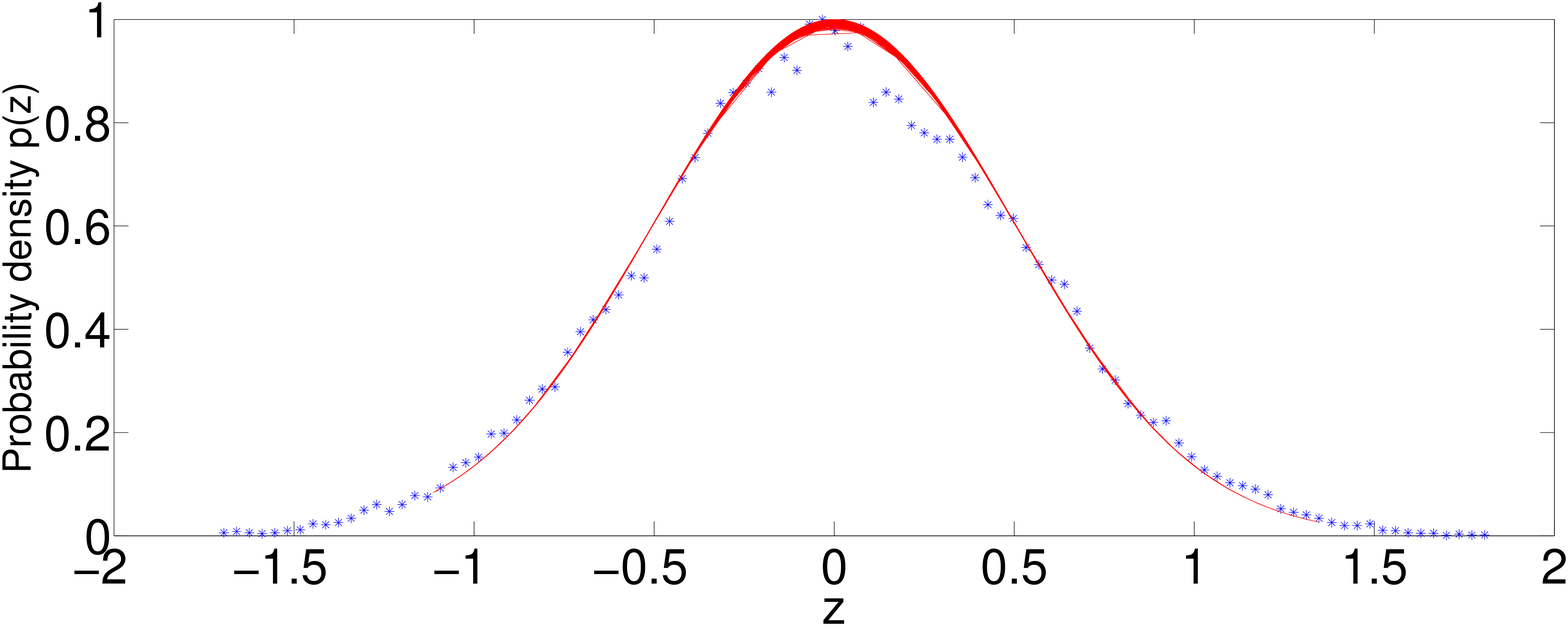}
\caption{Statistical properties obtained of the system (\ref{eq:modeloprop}). (a) shows a linear growth of the mean square displacement. Probability density obtained from the motion showed by normalized histogram approximation ( dotted blue  curve),  for  displacement (b),  velocity (c), and acceleration (d), compared with  theoretical Gaussian distribution ( dotted  red curve).} 
\label{fig:dis}
\end{figure}

\begin{figure}[h]
	\centering
		\includegraphics[width=0.70\textwidth]{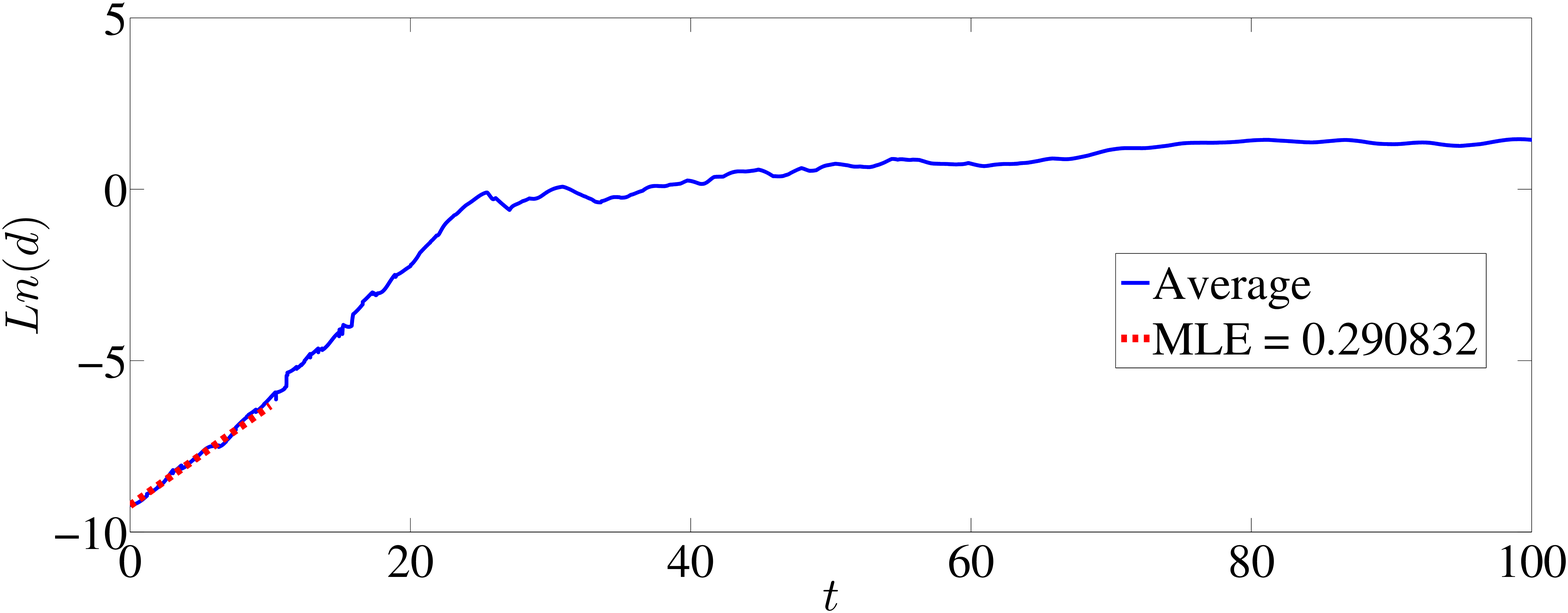}
	\caption{Maximum Lyapunov exponent obtained of the proposed system (\ref{eq:modeloprop}) with $\alpha=0.95$.}
	\label{fig:lyap}
\end{figure}

Figure \ref{fig:dis} shows the statistical properties obtained for the  time series of the system  (\ref{eq:modeloprop}). Figure~\ref{fig:dis}~(a)  displays the linear growth in time of the mean square displacement predicted, regarding with the  traditional  Brownian motion zero-mean Gaussian probability distributions Figures~\ref{fig:dis}~(b),~(c),~(d) show the  particle probability distributions of  displacement, velocity and acceleration, respectively, in which one can see that  the obtained distributions of the motion generated by our system have a better Gaussian approximation  than the obtained with integer order in Ref.~\cite{huerta},



As it is known, strong sensitivity
to initial conditions is an essential characteristic inherent of chaos. In the Brownian motion case  also is essential the strong sensibility to initial conditions, and both can be  characterized
by positive leading Lyapunov exponent. Even though both behaviors can be characterized by Lyapunov exponent the dynamic is completely different; Brownian motion as noise do not form an attractor in the phase space, {\it i.e.}, they are unbounded trayectories. Brownian and noise trajectories tend to infinity, while the chaotic dynamic generate an attractor localized within a certain area of phase space, so chaotic trayectories are bounded. In figure \ref{fig:lyap} the maximum Lyapunov exponent obtained, according with Ref.~\cite{rosenstein}, of the proposed system is shown to confirm the noise behavior of Brownian motion.  

Finally,  the correlation property of the signals generated by \eqref{eq:modeloprop} is characterized  by the detrended fluctuation analysis  (DFA) which was developed by Peng {\it et al.} \cite{Peng}. So the DFA thecnique help us to ensure that the proposed system generates Brownian motion, we apply the DFA evaluation method to the time series obtained with the parameter values before used. The DFA is an important tool for the detection of long-range auto- correlations in time series with non-stationarities. The DFA
is based on the random walk theory  which consists on a scaling analysis. The main advantages of the DFA over many other methods are that it allows the detection
of long-range correlations of a signal embedded in seemingly nonstationary time series, and also avoids the spurious detection of apparent long-range correlations that are an artifact of non-stationarity.  

The DFA procedure consists on the next four steps

\begin{enumerate}
  \item [*]Compute the time series mean $\bar{x}$.
  \item  [*]The interbeat interval time series (of total length $N$) is first integrated
  \begin{equation}
	y(k)=\sum_{i=1}^k[x(i)-\bar{x}].
	\end{equation}
\item [*]The integrated time series is divided into boxes of equal
length $n$. the local trend is obteined by mean-squares and is removed to each box.
\item  [*]The rootmean-
square fluctuation of this integrated and detrended time
series is calculated by 
\begin{equation}
f(n)=\sqrt{\frac{1}{N}\sum_{k=1}^N[y(k)-y_n(k)]^2}.
\end{equation}
\item [*]The fluctuations
can be characterized by a scaling exponent $\eta$, the slope of
the line relating $log F(n)$ to $log~ n$
\begin{equation}
f_m(n)\sim n^\eta.
\end{equation}
\end{enumerate}  

When the scaling exponent $\eta > 0.5$ three distinct regimes
can be defined as follows.\\

\begin{enumerate}
\item [1]If $\eta \approx 1$, DFA defines $1/f$ noise.
\item [2]If  $\eta > 1$, DFA defines a non stationary or unbounded behavior.
\item [3]If  $\eta \approx 1.5$, DFA defines Brownian motion or Brownian noise.
\end{enumerate}

\begin{figure}[h]
	\centering
		\includegraphics[width=0.70\textwidth]{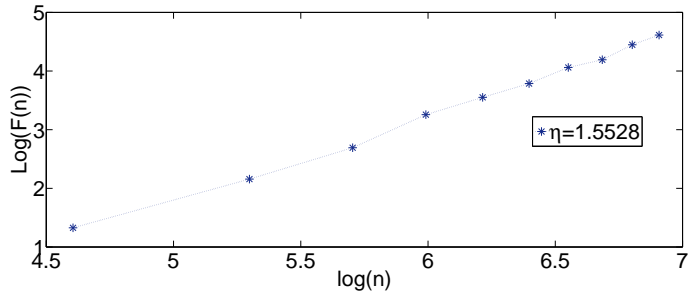}
	\caption{$\eta \approx 1:5$ obtained by DFA
indicates the Brownian behavior of the observed motion.}
	\label{fig:dfa}
\end{figure}

The scaling law with $\eta= 1:5401$ revealed by the DFA and shown in Figure \ref{fig:dfa} confirms the Brownian behavior.

\section{\label{sec:Conclusions}Concluding remarks}

A fractional deterministic model to generate Brownian motion has been presented by considering a fractional derivative order and the jerk equation instead of  the stochastic process and interger derivative order in the Langevin equation. These changes modify the Langevin equation by adding an additional degree
of freedom, so a three-dimensional model was obtained. By means of considering the fractional derivative order, the statistical properties were improved compared to their integer derivative order. The new variable introduced in the system defined by a third differential equation has a Gaussian probability density distribution which was confirmed with numerical simulations.   The statistic analysis of time series obtained with the proposed model  displayed typical characteristics of Brownian motion, namely, a linear growth of the mean square displacement, a Gaussian probability density distribution for displacement, velocity and acceleration. Furthermore, the Brownian behavior was confirmed by an approximately 1.5 power law scaling of the fluctuation. These results show that time series obtained of the proposal model fulfill the main characteristics of the  Brownian motion via stochastic process.

Based on  these results which were obtained by using unstable dissipative system, it  can be thought that the methodology presented in this work
could be used to construct models under external force fields or behaviors of anomalous diffusion. Additionally the development of adequate realistic models with real experimental time series in order to obtain the Brownian motion in real systems.

\section{Acknowledgements}
H.E.G.V is a doctoral fellow of the CONACYT in the Graduate Program on control and dynamical systems at DMAp-IPICYT.

\end{document}